\documentstyle[twocolumn]{article}
\newcommand{\cqg}[1]{{\em Class.\ Quan.\ Grav.\ }{\bf #1}}
\newcommand{\grg}[1]{{\em Gen.\ Rel.\ Grav.\ }{\bf #1}}
\newcommand{\np}[1]{{\em Nucl.\ Phys.\ }{\bf #1}}
\newcommand{\pr}[1]{{\em Phys.\ Rev.\ }{\bf #1}}
\newcommand{\prl}[1]{{\em Phys.\ Rev.\ Lett.\ }{\bf #1}}
\newcommand{\pl}[1]{{\em Phys.\ Lett.\ }{\bf #1}}
\newcommand{\jmp}[1]{{\em J. Math.\ Phys.\ }{\bf #1}}
\newcommand{\jgp}[1]{{\em J. Geom.\ Phys.\ }{\bf #1}}
\newcommand{\cmp}[1]{{\em Commun.\ Math.\ Phys.\ }{\bf #1}}
\newcommand{\mpl}[1]{{\em Mod.\ Phys.\ Lett.\ }{\bf #1}}
\newcommand{\ijmp}[1]{{\em Int.\ J. Mod.\ Phys.\ }{\bf #1}}
\newcommand{\apny}[1]{{\em Ann.\ Phys.\ (N.Y.) }{\bf #1}}

\begin{document}
\onecolumn

\title{Bibliography of Publications related to \\ Classical and Quantum Gravity
\\ in terms of the Ashtekar Variables}
\author{
       Last updated by \\
       B. Br\"ugmann \\
       Max-Planck-Institut f\"ur Physik, F\"ohringer Ring 6 \\
       80805 M\"unchen, Germany\\
       E-mail: bruegman@iws170.mppmu.mpg.de
}

\date{September 11, 1993}
\maketitle

\begin{abstract}
This bibliography attempts to give a comprehensive overview of all the
literature related to the Ashtekar variables. The original version was
compiled by Peter H\"ubner in 1989, and it has been subsequently
updated by Gabriela Gonzalez and Bernd Br\"ugmann. Information about
additional literature, new preprints, and especially corrections are
always welcome.
\end{abstract}

\newpage
\twocolumn

\section*{Pointers}

Given the number of references that follow, let me point out to the
non-specialist some of the references that can serve as entry points
into the literature.

First of all, for a complete and authorative presentation of canonical
gravity in the Ashtekar variables there is of course Ashtekar's latest
book [4] which appeared in 1991. The most recent general introduction
to the new variables by Ashtekar are his Les Houches lectures of 1992
[301].

Rather complete reviews of canonical gravity in the Ashtekar variables
can be found in Rovelli [171] and Kodama [323]. For a recent status
report see Smolin [346].  For a critical appraisal of canonical
quantum gravity see Kucha{\v r} [325]. An overview over different
approaches to quantum gravity is given in Isham [159].

Finally let me mention a few more specialized references. A clear
and detailed exposition of connection dynamics is given by Romano in
[8,342]. For newer developments related to matter couplings (geometric
approach) see Peld\'an [339]. The definition of the loop representation is
discussed in Br\"ugmann [11]. Pullin [340] gives an introduction to
results obtained via the loop representation in (unreduced) quantum
gravity.

\newpage

\section*{Books and Dissertations}

\begin{enumerate}

\item
Abhay Ashtekar and {invited contributors}.
 {\em New Perspectives in Canonical Gravity}.
 Lecture Notes.  Napoli, Italy: Bibliopolis, February 1988.
[Errata published as Syracuse University preprint by Joseph D. Romano
  and Ranjeet S. Tate.]

\item
V. Husain. {\it Investigations on the canonical quantization of gravity.}
Ph.D. Thesis, Yale University (1988).

\item
Paul. A. Renteln. {\em Non-perturbative approaches to Quantum Gravity. }
Ph.D. Thesis, Harvard University (1988).

\item
Abhay Ashtekar.
 {\em Lectures on non-perturbative canonical gravity.}
 (Notes prepared in collaboration with R. Tate).
Advanced Series in Astrophysics and Cosmology-Vol. 6.
Singapore: World Scientific, 1991.

\item
R. Capovilla.
{\em The self-dual spin connection as the fundamental gravitational
variable.}
Ph.D. Thesis, University of Maryland (1991).

\item
S. Koshti.
{\em Applications of the Ashtekar variables in Classical Relativity}.
Ph. D. Thesis, University of Poona (June 1991).

\item
D. Rayner.
{\em New variables in canonical quantisation and quantum gravity.}
Ph.D. Thesis, University of London (1991).

\item
J. D.  Romano.
 {\em Geometrodynamics vs. Connection Dynamics (in the context
of (2+1)- and (3+1)-gravity)}.
 Ph.D. Thesis, Syracuse University (1991), see also gr-qc/9303032

\item
C. Soo.
{\em Classical and quantum gravity with Ashtekar variables.}
Ph.D. Thesis, Virginia Polytechnic Institute and State University.
VPI-IHEP-92-11 (July 1992)

\item
R.S. Tate.
{\em An algebraic approach to the quantization of constrained systems:
finite dimensional examples.}
Ph.D. Thesis, Syracuse University (Aug. 1992), gr-qc/9304043

\item
B. Br\"ugmann.
{\em On the constraints of quantum general relativity in the loop
representation.}
Ph.D. Thesis, Syracuse University (May 1993)


\newpage

\section*{Papers}

\section*{1980}

\item
Paul Sommers.
 Space spinors.
 {\em J. Math. Phys.} {\bf 21}(10):2567--2571, October 1980.

\section*{1981}

\item
Amitabha Sen.
 On the existence of neutrino ``zero-modes'' in vacuum spacetimes.
 {\em J. Math. Phys.} {\bf 22}(8):1781--1786, August 1981.

\section*{1982}

\item
Abhay Ashtekar and G.T. Horowitz.
 On the canonical approach to quantum gravity.
 {\em Phys. Rev.} {\bf D26}:3342--3353, 1982.

\item
Amitabha Sen.
 Gravity as a spin system.
 {\em Phys. Lett. } {\bf B119}:89--91, December 1982.

\section*{1984}

\item
Abhay Ashtekar.
 On the {H}amiltonian of general relativity.
 {\em Physica} {\bf A124}:51--60, 1984.

\item
E.~T. Newman.
 Report of the workshop on classical and quantum alterate theories of
  gravity.
 In B.~Bertotti, F.~{de Felice}, and A.~Pascolini, editors, {\em The
  Proceedings of the 10th International Conference on General Relativity and
  Gravitation}, Amsterdam, 1984.

\section*{1986}

\item
Abhay Ashtekar.
 New variables for classical and quantum gravity.
 {\em Phys. Rev. Lett.} {\bf 57}(18):2244--2247, November 1986.

\item
Abhay Ashtekar.
 Self-duality and spinorial techniques in the canonical approach to
  quantum gravity.
 In C.~J. Isham and R.~Penrose, editors, {\em Quantum Concepts in
  Space and Time}, pages 303--317. Oxford University Press, 1986.

\item
Robert~M. Wald.
 Non-existence of dynamical perturbations of {S}chwarzschild with
  vanishing self-dual part.
 {\em Class. Quan. Grav.} {\bf 3}(1):55--63, January 1986.

\newpage

\section*{1987}

\item
Abhay Ashtekar.
 New {H}amiltonian formulation of general relativity.
 {\em Phys. Rev. } {\bf D36}(6):1587--1602, September 1987.

\item
Abhay Ashtekar.
 {E}instein constraints in the {Y}ang-{M}ills form.
 In G.~Longhi and L~Lusanna, editors, {\em Constraint's Theory and
  Relativistic Dynamics}, Singapore, 1987. World Scientific.

\item
Abhay Ashtekar, Pawel Mazur, and Charles~G. Torre.
 {BRST} structure of general relativity in terms of new variables.
 {\em Phys. Rev. } {\bf D36}(10):2955--2962, November 1987.

\item
John~L. Friedman and Ian Jack.
 Formal commutators of the gravitational constraints are not
  well-defined: A translation of {A}shtekar's ordering to the {S}chr{\"o}dinger
  representation.
 {\em Phys. Rev. } {\bf D37}(12):3495--3504, June 1987.

\item
Ted Jacobson and Lee Smolin.
 The left-handed spin connection as a variable for canonical gravity.
 {\em Phys. Lett. } {\bf B196}(1):39--42, September 1987.

\item
Joseph Samuel.
 A {L}agrangian basis for {A}shtekar's reformulation of canonical
  gravity.
 {\em Pram{\=a}na-J Phys.} {\bf 28}(4):L429-L432, April 1987.

\item
N.~C. Tsamis and R.~P. Woodard.
 The factor ordering problem must be regulated.
 {\em Phys. Rev.}  {\bf D36}(12):3641--3650, December 1987.

\newpage

\section*{1988}

\item
Abhay Ashtekar.
 A $3+1$ formulation of {E}instein self-duality.
 In J.~Isenberg, editor, {\em Mathematics and General Relativity},
  Providence, 1988. American Mathematical Society.

\item
Abhay Ashtekar.
 Microstructure of space-time in quantum gravity.
 In K.~C. Wali, editor, {\em Proceedings of the Eight Workshop in
  Grand Unification}, Singapore, 1988. World Scientific.

\item
Abhay Ashtekar.
 New perspectives in canonical quantum gravity.
 In B.~R. Iyer, A.~Kembhavi, J.~V. Narlikar, and C.~V. Vishveshwara,
  editors, {\em Highlights in Gravitation and Cosmology}. Cambridge University
  Press, 1988.

\item
Abhay Ashtekar, Ted Jacobson, and Lee Smolin.
 A new characterization of half-flat solutions to {E}instein's
  equation.
 {\em Commun. Math. Phys.} {\bf 115}:631--648, 1988.

\item
Ingemar Bengtsson.
 Note on {A}shtekar's variables in the spherically symmetric case.
 {\em Class. Quan. Grav.} {\bf 5}(10):L139--L142, October 1988.

\item
R. Gianvittorio, R. Gambini and A. Trias.
\pr{D38} (1988) 702

\item
J.~N. Goldberg.
 A {H}amiltonian approach to the strong gravity limit.
 {\em Gen. Rel. Grav.} {\bf 20}(9):881--891, September 1988.

\item
J.~N. Goldberg.
 Triad approach to the {H}amiltonian of general relativity.
 {\em Phys. Rev. } {\bf D37}(8):2116--2120, April 1988.

\item
Viqar Husain.
 The {$G_{\mbox{Newton}}\rightarrow\infty$} limit of quantum gravity.
 {\em Class. Quan. Grav.} {\bf 5}(4):575--582, April 1988.

\item
Ted Jacobson.
 Fermions in canonical gravity.
 {\em Class. Quan. Grav.} {\bf 5}(10):L143--L148, October 1988.

\item
Ted Jacobson.
 New variables for canonical supergravity.
 {\em Class. Quan. Grav.} {\bf 5}:923--935, 1988.

\item
Ted Jacobson.
 Superspace in the self-dual representation of quantum gravity.
 In J.~Isenberg, editor, {\em Mathematics and General Relativity},
  Providence, 1988. American Mathematical Society.

\item
Ted Jacobson and Lee Smolin.
 Covariant action for {A}shtekar's form of canonical gravity.
 {\em Class. Quan. Grav.} {\bf 5}(4):583--594, April 1988.

\item
Ted Jacobson and Lee Smolin.
 Nonperturbative quantum geometries.
 {\em Nucl. Phys.} {\bf B299}(2):295--345, April 1988.

\item
Hideo Kodama.
 Specialization of {A}shtekar's formalism to {B}ianchi cosmology.
 {\em Prog. Theor. Phys.} {\bf 80}(6):1024--1040, December 1988.

\item
Carlo Rovelli and Lee Smolin.
 Knot theory and quantum gravity.
 {\em Phys. Rev. Lett.} {\bf 61}:1155--1158, 1988.

\item
Joseph Samuel.
 Gravitational instantons from the {A}shtekar variables.
 {\em Class. Quan. Grav.} {\bf 5}:L123--L125, 1988.

\item
Lee Smolin.
 Quantum gravity in the self-dual representation.
 In J.~Isenberg, editor, {\em Mathematics and General Relativity},
  Providence, 1988. American Mathematical Society.

\item
C.~G. Torre.
 The propagation amplitude in spinorial gravity.
 {\em Class. Quan. Grav.} {\bf 5}:L63--L68, 1988.

\item
Edward Witten.
 (2+1) dimensional gravity as an exactly soluble system.
 {\em Nucl. Phys.} {\bf B311}(1):46--78, December 1988.

\newpage

\section*{1989}

\item
Abhay Ashtekar.
 Non-pertubative quantum gravity: A status report.
 In M.~Cerdonio, R.~Cianci, M.~Francaviglia, and M.~Toller, editors,
  {\em General Relativity and Gravitation}. Singapore: World Scientific, 1989.

\item
Abhay Ashtekar.
 Recent developments in {H}amiltonian gravity.
 In B.~Simon, I.~M. Davies, and A.~Truman, editors, {\em The
  Proceedings of the {IX}th International Congress on Mathematical Physics},
Swansea UK, July 1988.(Bristol, UK: Adam Hilger, 1989).

\item
Abhay Ashtekar.
 Recent developments in quantum gravity.
 In E.~J. Fenyves, editor, {\em Proceedings of the Texas Symposium on
  Relativistic Astrophysics}. New York Academy of Science, 1989.

\item
Abhay Ashtekar.
 Recent Developments in Quantum Gravity. {\it Annals of the New York
Academy of Sciences} {\bf 571}, 16-26. December 1989.

\item
Abhay Ashtekar, A.~P. Balachandran, and S.~G. Jo.
 The {CP}-problem in quantum gravity.
 {\em Int. Journ. Theor. Phys.} {\bf A4}:1493--1514, 1989.

\item
Abhay Ashtekar, Viqar Husain, Carlo Rovelli, Joseph Samuel, and Lee Smolin.
 $2+1$ quantum gravity as a toy model for the $3+1$ theory.
 {\em Class. Quan. Grav.} {\bf 6}:L185--L193, 1989.

\item
Abhay Ashtekar and Joseph~D. Romano.
 {C}hern-{S}imons and {P}alatini actions and ($2+1$)-gravity.
 {\em Phys. Lett. } {\bf B229}(1,2):56--60, October 1989.

\item
Abhay Ashtekar, Joseph~D. Romano, and Ranjeet~S. Tate.
 New variables for gravity: Inclusion of matter.
 {\em Phys. Rev. } {\bf D40}(8):2572--2587, October 1989.

\item
Abhay Ashtekar and Joseph~D. Romano.
 Key ($3+1$)-equations in terms of new variables (for numerical
  relativity).
 Syracuse University Report (1989).

\item
Ingemar Bengtsson.
 {Y}ang-{M}ills theory and general relativity in three and four
  dimensions.
 {\em Phys. Lett. } {\bf B220}:51--53, 1989.

\item
Ingemar Bengtsson.
 Some remarks on space-time decomposition, and degenerate metrics, in
  general relativity.
 {\em Int. J.  Mod. Phys. } {\bf A4}(20):5527--5538,
  1989.

\item
Riccardo Capovilla, John Dell, and Ted Jacobson.
 General relativity without a metric.
 {\em Phys. Rev. Lett.} {\bf 63}(21):2325--2328, November 1989.

\item
Steven Carlip.
 Exact quantum scattering in 2+1 dimensional gravity.
 {\em Nucl. Phys.} {\bf B324}(1):106--122, 1989.

\item
B. P. Dolan.
 On the generating function for Ashtekar's canonical transformation.
 {\em Phys. Lett. } {\bf B233}(1,2):89-92 , December 1989.

\item
Tevian Dray, Ravi Kulkarni, and Joseph Samuel.
 Duality and conformal structure.
 {\em J. Math. Phys.} {\bf 30}(6):1306--1309, June 1989.

\item
N.~N. Gorobey and A.~S. Lukyanenko.
 The closure of the constraint algebra of complex self-dual gravity.
 {\em Class. Quan. Grav.} {\bf 6}(11):L233--L235, November 1989.

\item
M.~Henneaux, J.~E. Nelson, and C.~Schomblond.
 Derivation of {A}shtekar variables from tetrad gravity.
 {\em Phys. Rev. } {\bf D39}(2):434--437, January 1989.

\item
A. Herdegen. Canonical gravity from a variation principle in a copy of
a tangent bundle. {\it Class.  Quan. Grav.} {\bf 6}(8):1111-24,
(1989).

\item
G. T. Horowitz.
Exactly soluble diffeomorphism invariant theories. {\it Commun.
Math. Phys.} {\bf 125}(3): 417-37, 1989.

\item
Viqar Husain.
 Intersecting loop solutions of the {H}amiltonian constraint of
  quantum general relativity.
 {\em Nucl. Phys.} {\bf B313}:711--724, 1989.

\item
Viqar Husain and Lee Smolin.
 Exactly solvable quantum cosmologies from two {K}illing field
  reductions of general relativity.
 {\em Nucl. Phys.} {\bf B327}:205--238, 1989.

\item
V.~Khatsymovsky.
 Tetrad and self-dual formulation of {R}egge calculus.
 {\em Class. Quan. Grav.} {\bf 6}(12):L249--L255, December 1989.

\item
Sucheta Koshti and Naresh Dadhich.
 Degenerate spherical symmetric cosmological solutions using
  {A}shtekar's variables.
 {\em Class. Quan. Grav.} {\bf 6}:L223--L226, 1989.

\item
Stephen~P. Martin.
 Observables in 2+1 dimensional gravity.
 {\em Nucl. Phys.} {\bf 327}(1):78--204, November 1989.

\item
L.~J. Mason and E.~T. Newman.
 A connection between {E}instein and {Y}ang-{M}ills equations.
 {\em Commun. Math. Phys.} {\bf 121}(4):659--668, 1989.

\item
J.~E. Nelson and T.~Regge.
 Group manifold derivation of canonical theories.
 {\em Int. J. Mod. Phys.} {\bf A4},2021 (1989).

\item
Paul Renteln and Lee Smolin.
 A lattice approach to spinorial quantum gravity.
 {\em Class. Quan. Grav.} {\bf 6}:275--294, 1989.

\item
Amitabha Sen and Sharon Butler.
 The quantum loop.
 {\em The Sciences}: 32--36, November/December 1989.

\item
L. Smolin.
 Invariants of links and critical points of the {C}hern-{S}imon path
  integrals.
 {\em Mod. Phys. Lett.} {\bf A4}:1091--1112, 1989.

\item
L. Smolin.
Loop representation for quantum gravity in 2+1 dimensions.
In the {\em Proceedings of the John's Hopkins Conference on Knots,
Topology and Quantum Field Theory}, ed. L. Lusanna (World Scientific,
Singapore 1989)

\item
Sanjay~M. Wagh and Ravi~V. Saraykar.
 Conformally flat initial data for general relativity in {A}shtekar's
  variables.
 {\em Phys. Rev. } {\bf D39}(2):670--672, January 1989.

\item
Edward Witten.
 Gauge theories and integrable lattice models.
 {\em Nucl. Phys.} {\bf B322}(3):629--697, August 1989.

\item
Edward Witten.
 Topology-changing amplitudes in (2+1) dimensional gravity.
 {\em Nucl. Phys.} {\bf B323}(1):113--122, August 1989.

\newpage

\section*{1990}

\item
C. Aragone and A. Khouder .
 Vielbein gravity in the light-front gauge.
 {\em Class. Quan. Grav.} {\bf 7}:1291--1298, 1990.

\item
Abhay Ashtekar.
Old problems in the light of new variables.
In {\em Proceedings of the Osgood Hill Conference on Conceptual
Problems in Quantum Gravity}, eds. A. Ashtekar and J. Stachel
(Birkh\"auser, Boston 1991)

\item
Abhay Ashtekar.
 Self duality, quantum gravity, {W}ilson loops and all that.
 In N.~Ashby, D.~F. Bartlett, and W.~Wyss, editors, {\em Proceedings
  of the 12th International Conference on General Relativity and Gravitation}.
  Cambridge University Press, 1990.

\item
Abhay Ashtekar and Jorge Pullin.
 {B}ianchi cosmologies: A new description.
 {\em Proc. Phys. Soc. Israel} {\bf 9}:65-76 (1990).

\item
Abhay Ashtekar.
 Lessons from 2+1 dimensional quantum gravity.
In {\em "Strings 90"} edited
by R. Arnowitt et al (Singapore: World Scientific, 1990).

\item
Ingemar Bengtsson.
 A new phase for general relativity?
 {\em Class. Quan. Grav.} {\bf 7}(1):27--39, January 1990.

\item
Ingemar Bengtsson.
 P, T, and the cosmological constant.
 {\em Int. J.  Mod. Phys. } {\bf A5}(17):3449-3459 (1990).

\item
Ingemar Bengtsson.
 Self-Dual Yang-Mills fields and Ashtekar variables.
 {\em Class. Quan. Grav.} {\bf 7}:L223-L228 (1990)

\item
Ingemar Bengtsson and P. Peld{\' a}n.
Ashtekar variables, the theta-term, and the cosmological constant.
{\em Phys. Lett.} {\bf B244}(2): 261-64, 1990.

\item
M.~P. Blencowe.
 The {H}amiltonian constraint in quantum gravity.
 {\em Nuc. Phys.} {\bf B341}(1):213, 1990.

\item
L.~Bombelli and R.~J. Torrence.
 Perfect fluids and {A}shtekar variables, with applications to
  {K}antowski-{S}achs models.
 {\em Class.Quan. Grav.} {\bf 7}:1747 (1990).

\item
Riccardo Capovilla, John Dell, and Ted Jacobson.
 Gravitational instantons as {SU(2)} gauge fields.
 {\em Class.Quan. Grav.} {\bf 7}(1):L1--L3, January 1990.

\item
Steven Carlip.
 Observables, gauge invariance and time in 2+1 dimensional gravity.
 {\em Phys. Rev.} {\bf D42}, 2647-2654 (October 1990).

\item
S. Carlip and S. P. de Alwis.
 Wormholes in (2+1)-gravity.
 {\em Nuc. Phys.} {\bf B337}:681-694, June 1990.

\item
G. Chapline.
Superstrings and Quantum Gravity.
{\em Mod. Phys. Lett.}{\bf A5}:2165-72 (1990).

\item
R. Floreanini and R. Percacci.
 Canonical algebra of GL(4)-invariant gravity.
 {\em Class.Quan. Grav.} {\bf 7}:975--984, 1990.

\item
R. Floreanini and R. Percacci.
 Palatini formalism and new canonical variables for GL(4)-invariant
gravity.
{\em Class. Quan. Grav.} {\bf 7}: 1805-18, 1990.

\item
R. Floreanini and R. Percacci.
 Topological pregeometry.
 {\em Mod. Phys. Lett.} {\bf A5}: 2247-51, 1990.

\item
Takeshi Fukuyama and Kiyoshi Kaminura.
 Complex action and quantum gravity.
{\em Phys. Rev.} {\bf D41}:1105-11, February 1990.

\item
G.~Gonzalez and J.~Pullin.
 {BRST} quantization of 2+1 gravity.
 {\em Phys. Rev. } {\bf D42}(10): 3395-3400 (1990).
[Erratum: {\em Phys. Rev.} {\bf 43}: 2749, April 1991].

\item
N.~N. Gorobey and A.~S. Lukyanenko.
 The {A}shtekar complex canonical transformation for supergravity.
 {\em Class. Quan. Grav.} {\bf 7}(1):67--71, January 1990.

\item
C. Holm.
 Connections in Bergmann manifolds.
 {\em Int. Journ. Theor. Phys.} {\bf A29}(1):23-36, January 1990.

\item
V. Husain and K. Kucha{\v r}.
 General covariance, the New variables, and dynamics without dynamics.
{\em Phys. Rev.} {\bf D42}(12)4070-4077 (December 1990).

\item
Viqar Husain and Jorge Pullin.
 Quantum theory of space-times with one Killing field.
 {\em Modern Phys. Lett. } {\bf A5}(10):733-741, April 1990.

\item
K. Kamimura and T. Fukuyama. Ashtekar's formalism in 1st order tetrad form.
{\em Phys. Rev.}  {\bf D41}(6): 1885-88, 1990.

\item
H. Kodama.
 Holomorphic wavefunction of the universe.
 {\em Phys. Rev.} {\bf D42}: 2548-2565 (October 1990).

\item
Sucheta Koshti and Naresh Dadhich.
 On the self-duality of the {W}eyl tensor using {A}shtekar's
  variables.
 {\em Class. Quan. Grav.} {\bf 7}(1):L5--L7, January 1990.

\item
Noah Linden.
 New designs on space-time foams.
 {\em Physics World} {\bf 3}(3):30-31, March 1990.

\item
N.Manojlovic.
 Alternative loop variables for canonical gravity.
 {\em Class. Quan. Grav.} {\bf 7}:1633-1645. (1990).

\item
E.~W. Mielke.
 Generating functional for new variables in general relativity and
  {P}oincare gauge theory.
 {\em Phys. Lett.} {\bf A149}: 345-350 (1990).

\item
E.~W. Mielke.
 Positive gravitational energy proof from complex variables?
 {\em Phys. Rev.} {\bf D42}(10): 3338-3394 (1990).

\item
Peter Peld\'{a}n.
 Gravity coupled to matter without the metric.
{\em Phys. Lett.} {\bf B248}(1,2): 62-66 (1990).

\item
D.~Rayner.
 A formalism for quantising general relativity using non-local
  variables.
 {\em Class. Quan. Grav.} {\bf 7}(1):111--134, January 1990.

\item
D.~Rayner.
 {H}ermitian operators on quantum general relativity loop space.
 {\em Class. Quan. Grav.} {\bf 7}(4):651--661, April 1990.

\item
Paul Renteln.
 Some results of {SU}(2) spinorial lattice gravity.
 {\em Class. Quan. Grav.} {\bf 7}(3):493--502, March 1990.

\item
D.C. Robinson and C. Soteriou.
 Ashtekar's new variables and the vacuum constraint equations.
 {\em Class. Quan. Grav.} {\bf 7}(11): L247-L250 (1990).

\item
Carlo Rovelli and Lee Smolin.
 Loop representation of quantum general relativity.
 {\em Nuc. Phys.} {\bf B331}(1): 80-152, February 1990.

\item
M.~Seriu and H.~Kodama.
 New canonical formulation of the {E}instein theory.
 {\em Prog. Theor. Phys.} {\bf 83}(1):7-12, January 1990.

\item
Lee Smolin.
 Loop representation for quantum gravity in $2+1$ dimensions.
 In {\em Proceedings of the 12th John Hopkins Workshop: Topology and
  Quantum Field Theory} (Florence, Italy), 1990.

\item
C. G. Torre.
 Perturbations of gravitational instantons.
 {\em Phys. Rev.} {\bf D41}(12) : 3620-3621, June 1990.

\item
C.~G. Torre.
 A topological field theory of gravitational instantons.
 {\em Phys. Lett } {\bf B252}(2):242-246 (1990).

\item
C. G. Torre.
 On the linearization stability of the conformally
(anti)self dual  {E}instein equations,
 {\em J. Math. Phys.} {\bf 31}(12): 2983-2986 (1990).

\item
H. Waelbroeck.
 2+1 lattice gravity.
 {\em Class. Quan. Grav.} {\bf 7}(1): 751--769, January 1990.

\item
M. Waldrop.
Viewing the Universe as a Coat of Chain Mail.
{\em Science} {\bf 250}: 1510-1511 (1990).

\item
R. P. Wallner
 New variables in gravity theories.
 {\em Phys. Rev. } {\bf D42}(2):441-448 ,July  1990.

\item
R.S. Ward.
 The SU($\infty$) chiral model and self-dual vacuum spaces.
 {\em Class. Quan. Grav.} {\bf 7}: L217-L222 (1990).

\newpage

\section*{1991}

\item
V. Aldaya and J. Navarro-Salas.
New solutions of the hamiltonian and diffeomorphism constraints of quantum
gravity from a highest weight loop representation.
{\em Phys. Lett.} {\bf B259}: 249-55, April 1991.

\item
Abhay Ashtekar.
Old problems in the light of new variables.
In {\em Proceedings of the Osgood Hill Conference on Conceptual
Problems in Quantum Gravity}, eds. A. Ashtekar and J. Stachel
(Birkh\"auser, Boston 1991)

\item
Abhay Ashtekar.
The winding road to quantum gravity.
In {\em Proceedings of the Osgood Hill Conference on Conceptual
Problems in Quantum Gravity}, eds. A. Ashtekar and J. Stachel
(Birkh\"auser, Boston 1991)

\item
Abhay Ashtekar.
Canonical Quantum Gravity.
In {\em The Proceedings of the 1990 Banff Workshop on Gravitational
Physics}, edited by R. Mann (Singapore: World Scientific, 1991), and
in the {\em Proceedings of SILARG VIII Conference}, edited by M.
Rosenbaum and M. Ryan (Singapore: World Scientific 1991).

\item
A. Ashtekar, C. Rovelli and L. Smolin.
Self duality and quantization.
{\em J. Geometry and Physics} Penrose Festschrift issue (1991).

\item
A. Ashtekar, C. Rovelli and L. Smolin.
Gravitons and loops.
{\em Phys. Rev.}{\bf D44}(6):1740-55, 15 September 1991.

\item
A. Ashtekar and J. Samuel.
Bianchi cosmologies: the role of spatial topology.
\cqg{8} (1991) 2191--215

\item
I. Bengtsson.
 The cosmological constants.
{\em Phys. Lett.} {\bf B254}:55-60, 1991.

\item
I. Bengtsson.
Self-duality and the metric in a family of neighbours of Einstein's equations.
J. Math. Phys. 32 (Nov. 1991) 3158--61

\item
Peter G. Bergmann and Garrit Smith.
Complex phase spaces and complex gauge groups in general relativity.
 {\em Phys. Rev.} {\bf D43}:1157-61,  February 1991.

\item
L. Bombelli.
Unimodular relativity, general covariance, time, and the Ashtekar
variables.
In {\em Gravitation. A Banff Summer Institute}, eds.~R. Mann and P.
Wesson (World Scientific 1991) 221--32

\item
L. Bombelli, W.E. Couch and R.J.Torrence.
Time as spacetime four-volume and the Ashtekar variables.
\pr{D44} (15. Oct. 1991) 2589--92

\item
B. Br\"{u}gmann.
 The method of loops applied to lattice gauge theory.
{\em Phys. Rev. } {\bf D43}: 566-79, January 1991.

\item
B. Br\"{u}gmann and J. Pullin.
 Intersecting N loop solutions of the Hamiltonian constraint
of Quantum Gravity.
 {\em Nuc. Phys.} {\bf B363}: 221-44, September 1991.

\item
R. Capovilla, J. Dell, T. Jacobson and L. Mason.
 Self dual forms and gravity.
{\em Class. Quan. Grav.} {\bf 8}: 41-57, January 1991.

\item
R. Capovilla, J. Dell and T. Jacobson.
 A pure spin-connection formulation of gravity.
{\em Class. Quan. Grav.} {\bf 8}: 59-74, January 1991.

\item
Steven Carlip.
 Measuring the metric in 2+1 dimensional quantum gravity.
{\em Class. Quan. Grav.} {\bf 8}:5-17, January 1991.

\item
S. Carlip and J. Gegenberg.
Gravitating topological matter in 2+1 dimensions.
{\em Phys. Rev.}{\bf D44}(2):424-28, 15 July 1991.

\item
L. Crane.
2-d physics and 3-d topology.
{\em Commun. Math. Phys.} {\bf 135}: 615-640, January 1991.

\item
N. Dadhich, S. Koshti and A. Kshirsagar.
 On constraints of pure connection formulation of General Relativity
for non-zero cosmological constant.
{\em Class. Quan. Grav.} {\bf 8}: L61-L64, March 1991.

\item
B.~P. Dolan.
 The extension of chiral gravity to {SL}(2,{C}).
In {\em Proceedings of the
1990 Banff Summer School on gravitation}, ed. by R. Mann (World Scientific,
Singapore 1991)

\item
R. Floreanini and R. Percacci.
 GL(3) invariant gravity without metric.
 {\em Class. Quan. Grav.}{\bf 8}(2):273-78, February 1991.

\item
G. Fodor and Z. Perjes.
Ashtekar variables without hypersurfaces.
{\em Proc. of Fifth Sem. Quantum Gravity, Moscow} (Singapore: World
Scientific 1991) 183--7

\item
T. Fukuyama and K. Kamimura.
Schwarzschild solution in Ashtekar formalism.
\mpl{A6} (1991) 1437--42

\item
R. Gambini.
Loop space representation of quantum general relativity and the group of loops.
{\em Phys. Lett.} {\bf B255}:180-88, February 1991.

\item
J.N. Goldberg.
Self-dual Maxwell field on a null cone.
\grg{23} (December 1991) 1403--1413

\item
J.N. Goldberg, E.T. Newman, and C. Rovelli.
On Hamiltonian systems with first class constraints.
\jmp{32}(10) (1991) 2739--43

\item
J. Goldberg, D.C. Robinson and C. Soteriou.
Null surface canonical formalism. In {\em  Gravitation and Modern Cosmology},
ed. Zichichi (Plenum Press, New York, 1991)

\item
J. Goldberg, D.C.Robinson and C. Soteriou.
A canonical formalism with a self-dual Maxwell field on a null surface.
In {\em 9th Italian Conference on General Relativity and Gravitational
Physics (P.G.  Bergmann Festschrift)}, ed. R. Cianci et al (World
Scientific, Singapore 1991)

\item
G.T. Horowitz.
Topology change in classical and quantum gravity.
{\em Class. Quan. Grav.} {\bf 8}:587-601, April 1991.

\item
V. Husain.
Topological quantum mechanics.
{\em Phys. Rev.} {\bf D43}: 1803-07, March 1991.

\item
C. J. Isham.
Loop Algebras and Canonical Quantum Gravity.
In Contemporary Mathematics, eds. by M. Gotay, V. Moncrief
and J. Marsden (American Mathematical Society, Providence, 1991).

\item
C. J. Isham.
Conceptual and geometrical problems in quantum gravity.
In {\em Recent aspects of quantum fields}, eds. H. Mitter and H. Gausterer
(Springer 1991) 123--229

\item
K. Kamimura, S. Makita and T. Fukuyama .
Spherically symmetric vacuum solution in Ashtekar's formulation of gravity.
\mpl{A6} (30. Oct. 1991) 3047--53

\item
C. Kozameh and E.T. Newman.
The O(3,1) Yang-mills equations and the Einstein equations.
{\em Gen. Rel. Grav.} {\bf 23}:87-98, January 1991.

\item
H. C. Lee and Z. Y. Zhu.
Quantum holonomy and link invariants.
{\em Phys. Rev.}{\bf D44}(4):R942-45, 15 August 1991.

\item
R. Loll.
 A new quantum representation for canonical gravity and SU(2)
Yang-Mills theory.
\np{B350} (1991) 831--60

\item
E. Mielke, F. Hehl.
Comment on ``General relativity without the metric''.
\prl{67} (Sept.\ 1991) 1370

\item
V. Moncrief and M. P. Ryan.
Amplitude-real-phase exact solutions for quantum mixmaster universes.
(to appear in {\em Phys. Rev.}{\bf D},1991).

\item
C.~Nayak.
 The loop space representation of 2+1 quantum gravity: physical
  observables,variational principles,  and the issue of time.
{\em Gen. Rel. Grav.} {\bf 23}: 661-70, June 1991.

\item
C.~Nayak.
Einstein-Maxwell theory in 2+1 dimensions.
{\em Gen. Rel. Grav.}{\bf 23}:981-90, September 1991.

\item
H. Nicolai.
The canonical structure of maximally extended supergravity in three dimensions.
\np{B353} (April 1991) 493

\item
P. Peld{\' a}n.
Legendre transforms in Ashtekar's theory of gravity.
\cqg{8} (Oct. 1991) 1765--83

\item
P. Peld{\' a}n.
Non-uniqueness of the ADM Hamiltonian for gravity.
\cqg{8} (Nov. 1991) L223--7

\item
C. Rovelli.
Ashtekar's formulation of general relativity and loop-space non-perturbative
quantum gravity : a report.
{\em Class. Quan. Grav.}{\bf 8}(9): 1613-1675, September 1991.

\item
Carlo Rovelli.
 Holonomies and loop representation in quantum gravity.
In {\em The Newman Festschrift}, ed. by A. Janis and J. Porter.
(Birkh{\" a}user, Boston 1991)

\item
Joseph Samuel.
 Self-duality in Classical Gravity.
In {\em The Newman Festschrift}, ed. by A. Janis and J. Porter.
(Birkh{\" a}user, Boston 1991)

\item
Lee Smolin.
 Nonperturbative quantum gravity via the loop representation.
In {\em Proceedings of the Osgood Hill Conference on Conceptual
Problems in Quantum Gravity}, eds. A. Ashtekar and J. Stachel
(Birkh\"auser, Boston 1991)

\item
S. Uehara.
A note on gravitational and SU(2) instantons with Ashtekar variables.
\cqg{8} (Nov. 1991) L229--34

\item
M. Varadarajan.
Non-singular degenerate negative energy solution to the Ashtekar equations.
\cqg{8} (Nov. 1991) L235--40

\item
K. Yamagishi and G.F. Chapline.
Induced 4-d self-dual quantum gravity: $\hat{W}_{\infty}$ algebraic approach.
{\em Class. Quan. Grav.}{\bf 8}(3):427-46, March 1991.

\item
J. Zegwaard.
Representations of quantum general relativity using Ashtekar's variables.
{\em Class. Quan. Grav.}{\bf 8} (July 1991) 1327--37

\newpage

\section*{1992}

\item
A. Ashtekar.
Loops, gauge fields and gravity.
In {\em Proceedings of the VIth Marcel Grossmann meeting on general
relativity}, eds.\ H. Sato and T. Nakamura (World Scientific, 1992),
and in {\em Proceedings of the VIIIth Canadian conference on
general relativity and gravitation}, edited by G. Kunstater et al
(World Scientific, Singapore 1992)

\item
A. Ashtekar and C. Isham.
Representations of the holonomy algebras of gravity and non-abelian
gauge theories.
\cqg{9} (June 1992) 1433--85

\item
A. Ashtekar and C. Isham.
Inequivalent observable algebras: a new ambiguity in field
quantisation.
\pl{B274} (1992) 393--398

\item
A. Ashtekar and J.D. Romano.
Spatial infinity as a boundary of space-time.
\cqg{9} (April 1992) 1069--100

\item
A. Ashtekar and C. Rovelli.
Connections, loops and quantum general relativity.
\cqg{9} suppl. (1992) S3--12

\item
A. Ashtekar and C. Rovelli.
A loop representation for the quantum Maxwell field.
\cqg{9} (May 1992) 1121--50

\item
A. Ashtekar, C. Rovelli and L. Smolin.
Self duality and quantization.
{\em J. Geom. Phys.} {\bf 8} (1992) 7--27

\item
A. Ashtekar, C. Rovelli and L. Smolin.
Weaving a classical geometry with quantum threads.
\prl{69} (1992) 237--40

\item
I. Bengtsson and O. Bostr\"om.
Infinitely many cosmological constants.
\cqg{9} (April 1992) L47--51

\item
I. Bengtsson and P. Peldan.
Another `cosmological' constant.
\ijmp{A7} (10 March 1992) 1287--308

\item
O. Bostr\"om.
Degeneracy in loop variables; some further results.
\cqg{9} (Aug. 1992) L83--86

\item
B. Br\"{u}gmann, R. Gambini and J. Pullin.
Knot invariants as nondegenerate quantum geometries.
\prl{68} (27 Jan. 1992) 431--4

\item
B. Br\"{u}gmann, R. Gambini and J. Pullin.
Knot invariants as nondegenerate states of four-dimensional quantum gravity.
In {\em Proceedings of the Twentieth International conference on
Differential Geometric Methods in Theoretical Physics, Baruch College, City
University of New York,1-7 June 1991}, S. Catto, A. Rocha eds. (World
Scientific, Singapore 1992)

\item
B. Br\"{u}gmann, R. Gambini and J. Pullin.
Jones polynomials for intersecting knots as physical states of quantum
gravity.
\np{B385} (Oct.\ 1992) 587--603

\item
L. N. Chang and C. P. Soo.
Ashtekar's Variables and the Topological Phase of Quantum Gravity.
In {\em Proceedings of the Twentieth International conference on
Differential Geometric Methods in Theoretical Physics, Baruch College, City
University of New York,1-7 June 1991}, S. Catto, A. Rocha eds. (World
Scientific, Singapore 1992)

\item
L. N. Chang and C. P. Soo.
BRST cohomology and invariants of four-dimensional gravity in
Ashtekar's variables.
\pr{D46} (Nov. 1992) 4257--62

\item
S. Carlip.
(2+1)-dimensional Chern-Simons gravity as a Dirac square root.
\pr{D45} (1992) 3584--90

\item
G. F\"ul\"op.
Transformations and BRST-charges in 2+1 dimensional gravitation.
gr-qc/9209003, \mpl{A7} (1992) 3495--3502

\item
T. Fukuyama.
Exact Solutions in Ashtekar Formalism.
In {\em Proceedings of the VIth Marcel Grossmann meeting on
general relativity}, eds.\ H. Sato and T. Nakamura (World Scientific, 1992)

\item
T. Fukuyama, K. Kamimura and S. Makita.
Metric from non-metric action of gravity.
\ijmp{D1} (1992) 363--70

\item
A. Giannopoulos and V. Daftardar.
The direct evaluation of the Ashtekar variables for any given metric
using the algebraic computing system STENSOR.
\cqg{9} (July 1992) 1813--22

\item
J. Goldberg.
Quantized self-dual Maxwell field on a null surface.
\jgp{8} (1992) 163--172

\item
J. Goldberg.
Ashtekar variables on null surfaces.
In {\em Proceedings of the VIth Marcel Grossmann meeting on
general relativity}, eds.\ H. Sato and T. Nakamura (World Scientific, 1992)

\item
J.N. Goldberg, J. Lewandowski, and C. Stornaiolo.
Degeneracy in loop variables.
\cmp{148} (1992) 377--402

\item
J.N. Goldberg, D.C. Robinson and C. Soteriou.
Null hypersurfaces and new variables.
\cqg{9} (May 1992) 1309--28

\item
J. Horgan.
Gravity quantized?
{\em Scientific American} (Sept.\ 1992) 18--20

\item
V. Husain.
2+1 gravity without dynamics.
\cqg{9} (March 1992) L33--36

\item
T. Jacobson and J.D. Romano.
Degenerate Extensions of general relativity.
\cqg{9} (Sept. 1992) L119--24

\item
C. Kim, T. Shimizu and K. Yushida.
2+1 gravity with spinor field.
\cqg{9} (1992) 1211-16

\item
S. Koshti.
Massless Einstein Klein-Gordon equations in the spin connection formulation.
\cqg{9} (1992) 1937--42

\item
J. Lewandowski.
Reduced holonomy group and Einstein's equations with a cosmological
constant.
\cqg{9} (Oct. 1992) L147--51

\item
R. Loll.
Independent SU(2)-loop variables and the reduced configuration space
of SU(2)-lattice gauge theory.
\np{B368} (1992) 121--42

\item
R. Loll.
Loop approaches to gauge field theory.
Syracuse SU-GP-92/6-2, in {\em Memorial
Volume for M.K. Polivanov, Teor. Mat. Fiz.} {\bf 91} (1992)

\item
A. Magnon.
Ashtekar variables and unification of gravitational and
electromagnetic interactions.
\cqg{9} suppl. (1992) S169--81

\item
J. Maluf.
Self-dual connections, torsion and Ashtekar's variables.
\jmp{33} (Aug.\ 1992) 2849--54

\item
N. Manojlovi{\' c} and A. Mikovi{\' c}.
Gauge fixing and independent canonical variables in the Ashtekar formalism
of general relativity.
\np{B382} (June 1992) 148--70

\item
N. Manojlovi{\' c} and A. Mikovi{\' c}.
Ashtekar Formulation of (2+1)-gravity on a torus.
\np{B385} (July 1992) 571--586

\item
E.W. Mielke.
Ashtekar's complex variables in general relativity and its teleparallelism
equivalent.
\apny{219} (1992) 78--108

\item
E.T. Newman and C. Rovelli.
Generalized lines of force as the gauge-invariant degrees of freedom
for general relativity and Yang-Mills theory.
\prl{69} (1992) 1300--3

\item
P. Peld\'an.
Connection formulation of (2+1)-dimensional Einstein gravity and
topologically massive gravity.
\cqg{9} (Sept. 1992) 2079--92

\item
L. Smolin.
The ${\rm G_{Newton}\rightarrow 0}$ limit of Euclidean quantum gravity.
\cqg{9} (April 1992) 883--93

\item
L. Smolin.
Recent developments in nonperturbative quantum gravity.
In {\em Proceedings of the XXII Gift International Seminar
on Theoretical Physics, Quantum Gravity and Cosmology, June 1991,
Catalonia, Spain} (World Scientific, Singapore 1992)

\item
V. Soloviev.
Surface terms in Poincare algebra in Ashtekar's formalism.
In {\em Proceedings of the VIth Marcel Grossmann meeting on
general relativity}, eds.\ H. Sato and T. Nakamura (World Scientific, 1992)

\item
Vladimir Soloviev.
How canonical are Ashtekar's variables?
\pl{B292} (???) 30--?

\item
R.S. Tate.
Polynomial constraints for general relativity using real
geometrodynamical variables.
\cqg{9} (Jan. 1992) 101--19

\item
C.G. Torre.
Covariant phase space formulation of parametrized field theory.
\jmp{33} (Nov. 1992) 3802--12

\item
R.P. Wallner.
Ashtekar's variables reexamined.
\pr{D46} (Nov. 1992) 4263--4285

\item
J. Zegwaard.
Gravitons in loop quantum gravity.
\np{B378} (July 1992) 288--308

\newpage

\section*{1993}

\item
A. Ashtekar.
Recent developments in classical and quantum theories of connections
including general relativity.
In {\em Advances in Gravitation and Cosmology}, eds.\ B. Iyer, A.
Prasanna, R. Varma and C. Vishveshwara (Wiley Eastern, New Delhi 1993)

\item
A. Ashtekar, lecture notes by R.S. Tate.
Physics in loop space.
In {\em Quantum gravity, gravitational radiation and large scale
structure in the universe}, eds. B.R.\ Iyer, S.V. Dhurandhar and K.
Babu Joseph (1993)

\item
A. Ashtekar and J. Lewandowski.
Completeness of Wilson loop functionals on the moduli space of
$SL(2,C)$ and $SU(1,1)$-connections.
gr-qc/9304044, \cqg{10} (June 1993) L69--74

\item
J.C. Baez.
Quantum gravity and the algebra of tangles.
\cqg{10} (April 1993) 673--94

\item
I. Bengtsson.
Some observations on degenerate metrics.
\grg{25} (Jan. 1993) 101--12

\item
J. Birman.
New points of view in knot theory.
{\em Bull. AMS}{\bf 28} (April 1993) 253--287

\item
B. Br\"ugmann, R. Gambini and J. Pullin.
How the Jones polynomial gives rise to physical states of quantum
general relativity.
\grg{25} (Jan.\ 1993) 1--6

\item
B. Br\"{u}gmann and J. Pullin.
On the constraints of quantum gravity in the loop representation.
\np{B390} (Feb.\ 1993) 399--438

\item
R. Capovilla and Jerzy Pleba\'nski.
Some exact solutions of the Einstein field equations in terms of the
self-dual spin connection.
\jmp{34} (Jan.\ 1993) 130--138

\item
Rodolfo Gambini and Jorge Pullin.
Quantum Einstein-Maxwell fields: a unified viewpoint from the loop
representation.
hep-th/9210110, \pr{D47} (June 1993) R5214--8

\item
V. Husain.
Ashtekar variables, self-dual metrics and $W_\infty$.
\cqg{10} (March 1993) 543--50

\item
V. Husain.
General covariance, loops, and matter.
gr-qc/9304010, \pr{D47} (June 1993) 5394--9

\item
C. Kiefer.
Topology, decoherence, and semiclassical gravity.
gr-qc/9306016, \pr{D47} (June 1993) 5413--21

\item
H.Y. Lee, A. Nakamichi and T. Ueno.
Topological two-form gravity in four dimensions.
\pr{D47} (Feb.\ 1993) 1563--68

\item
J. Lewandowski.
Group of loops, holonomy maps, path bundle and path connection.
\cqg{10} (1993) 879--904

\item
R. Loll.
Lattice gauge theory in terms of independent Wilson loops.
In {\em Lattice 92}, eds J. Smit and P. van Baal, \np{B}
(Proc.\ Suppl.) {\bf 30} (March 1993)

\item
R. Loll.
Yang-Mills theory without Mandelstam constraints.
\np{B400} (1993) 126--44

\item
R. Loll.
Loop variable inequalities in gravity and gauge theory.
\cqg{10} (Aug.\ 1993) 1471--1476

\item
J. Maluf.
Degenerate triads and reality conditions in canonical gravity.
\cqg{10} (April 1993) 805--9

\item
N. Manojlovi{\' c} and A. Mikovi{\' c}.
Canonical analysis of Bianchi models in the Ashtekar formulation.
\cqg{10} (March 1993) 559--74

\item
Peter Peld\'an.
Unification of gravity and Yang-Mills theory in 2+1 dimensions.
\np{B395} (1993) 239--62

\item
A. Rendall.
Comment on a paper of Ashtekar and Isham.
\cqg{10} (March 1993) 605--8

\item
J. Romano.
Geometrodynamics vs. connection dynamics.
gr-qc/9303032, \grg{25} (Aug.\ 1993) 759--854

\item
C. Rovelli.
Area is length of Ashtekar's triad field.
\pr{D47} (Feb.\ 1993) 1703--5

\item
R.S. Tate.
Constrained systems and quantization.
In {\em Quantum gravity, gravitational radiation and large scale
structure in the universe}, eds. B.R.\ Iyer, S.V. Dhurandhar and K.
Babu Joseph (1993)

\item J. Zegwaard.
The weaving of curved geometries.
\pl{B300} (Feb. 1993) 217--222

\newpage

\section*{Preprints older than 12 months}

\item
A. Ashtekar and J. Lee.
Weak field limit of general relativity: a new Hamiltonian formulation.
Syracuse preprint.

\item
A. Ashtekar, R.S. Tate and C. Uggla.
Minisuperspaces: observables, quantization and singularities.
SU-GP-92/2-6

\item
J.C. Baez.
Link invariants of finite type and perturbation theory.
UCR preprint (July 1992)

\item
C. Di Bartolo, R. Gambini, J. Griego, and L. Leal.
Loop space coordinates, linear representations of the diffeomorphism group
and knot invariants. Montevideo/Caracas IFFI-92-01

\item
I. Bengtsson.
Ashtekar's variables.
Goteborg-88-46 preprint (November 1988).

\item
I. Bengtsson.
Yang-Mills theory and general relativity in three dimensions and four
dimensions. Goteborg-89-1 preprint (January 1989).

\item
Ingemar Bengtsson.
 Complex actions, real slices, and {A}shtekar's variables.
 S-412 96 Preprint. February 1989, Inst. of Theor. Physics,
G\"{o}teburg, Sweden.

\item
I. Bengtsson.
Reality conditions for Ashtekar's variables.
Goteborg-89-4a preprint (October 1989).

\item
I. Bengtsson.
Self-dual Yang Mills fields and Ashtekar's variables.
Goteborg-90-11 preprint (April 1990).

\item
I. Bengtsson.
Degenerate metrics and an empty black hole.
Goteborg-90-45 (December 1990).

\item
I. Bengtsson.
Curvature tensors in an exact solution of Capovilla's equations.
Goteborg-91-5 (February 1991).

\item
I. Bengtsson.
Ashtekar's variables and the cosmological constant.
Goteborg preprint, 1991.

\item
Ola Bostr\"om.
Some new results about the cosmological constants.
G\"oteborg preprint ITP91-34

\item
Greorgy Burnett, Joseph~D. Romano, and Ranjeet~S. Tate.
 Polynomial coupling of matter to gravity using {A}shtekar variables.
Syracuse preprint.

\item
R. Capovilla.
 Generally covariant gauge theories.
 UMDGR 90-253 Preprint, May 1990.

\item
R. Capovilla and T. Jacobson.
Remarks on pure spin connection formulation of gravity.
Maryland preprint UMDGR-91-134

\item
Steven Carlip.
 2+1 dimensional quantum gravity and the Braid group.
 Talk given at the Workshop on Physics, Braids and Links, Banff Summer
  School in Theoretical Physics, August 1989.

\item
L. Chang and C. Soo.
Einstein manifolds in Ashtekar variables: explicit examples.
hep-th/9207056

\item
L. Crane.
Categorical physics.
Preprint ???.

\item
G. Esposito.
 Mathematical structures of space-time.
 Cambridge preprint DAMTP-R-gols, to appear in Fortschritte der Physik.

\item
R. Floreanini, R. Percacci and E. Spallucci.
Why is the metric non-degenerate?
SISSA 132/90/EP preprint (October 1990).

\item
R. Floreanini and R. Percacci.
Topological GL(3) invariant gravity.
SISSA-97-90-EP preprint (July 1990).

\item
H. Fort and R. Gambini.
Lattice QED with light fermions in the P representation.
IFFI preprint, 90-08

\item
Kazuo Ghoroku.
 New variable formalism of higher derivative gravity.

\item
B.~Grossmann.
 General relativity and a non-topological phase of topological
  {Y}ang-{M}ills theory.
 Inst. for Advanced Studies, Princeton, 1990 preprint.

\item
G. Harnett.
Metrics and dual operators.
Florida Atlantic University preprint, 1991.

\item
S. Hacyan.
Hamiltonian formulation of general relativity in terms of Dirac spinors.
UNAM Mexico preprint, 1991.

\item
G. Horowitz.
Ashtekar's approach to quantum gravity.
University of California preprint, 1991.

\item
G. t'Hooft.
A chiral alternative to the vierbein field in general relativity.
Uthrecht, THU-90/28 preprint (1990).

\item
H. Ikemori.
Introduction to two form gravity and Ashtekar formalism.
YITP-K-922 preprint (March 1991).

\item
W. Kalau.
Ashtekar formalism with real variables.
U. Of Wuppertal NIKHEF-H/91-03 Amsterdam preprint (December 1990).

\item
K. Kamimura and T. Fukuyama.
Massive analogue of Ashtekar-CDJ action.
gr-qc/9208010

\item
A. Kheyfets and W. A. Miller.
E. Cartan's moment of rotation in Ashtekar's theory of gravity.
Los Alamos preprint LA-UR-91-2605 (1991).

\item
S. Koshti and N. Dadhich.
 Gravitational instantons with matter sources using
Ashtekar variables.
 Inter Univ. Centre for Astron. and Astrophysics, Pune, India.
June 1990 preprint.

\item
A.M.R. Magnon.
Self duality and CP violation in gravity.
Univ. Blaise Pascal (France) preprint (1990).

\item
J. Maluf.
Symmetry properties of Ashtekar's formulation of canonical gravity.
Universidade de Brasilia preprint, 1991.

\item
J. Maluf.
Fermi coordinates and reference frames in the ECSK theory.
SU-GP-92/1-2

\item
L.~J. Mason and J{\"o}rg Frauendiener.
 The {S}parling 3-form, {A}shtekar variables and quasi-local mass,
  1989 preprint.

\item
N. O'Murchadha and M. Vandyck.
 Gravitational degrees of freedom in Ashtekar's formulation of
General Relativity.
 Univ. of Cork preprint - 1990

\item
Paul Renteln.
 Some notes on spinorial quantum gravity.
 Preprint.

\item
Carlo Rovelli and Lee Smolin.
 Loop representation for lattice gauge theory.
 1990 Pittsburgh and Syracuse preprint.

\item
C. Rovelli and L. Smolin.
Finiteness of diffeomorphism invariant operators in nonperturbative
quantum gravity. Syracuse University preprint SU-GP-91/8-1, August 1991.

\item
Lee Smolin.
The Problem of Quantum Gravity: a status report (Address to the AAAS meeting,
Washington D.C., February 1991). Syracuse preprint SU-GP-91/2-1.

\item
L. Smolin and M. Varadarajan.
Degenerate solutions and the instability of the perturbative vacuum in
nonperturbative formulations of quantum gravity.
Syracuse University preprint SU-GP-91/8-3, August 1991.

\item
T. Thiemann and H.A. Kastrup.
Canonical quantization of spherically symmetric gravity in Ashtekar's
self-dual representation.
\np{B399} (June 1993) 211--58

\item
C. G. Torre.
A deformation theory of self-dual Einstein spaces.
SU-GP-91/8-7, Syracuse University preprint, 1991.

\item
R. P. Wallner.
 A new form of Einstein's equations.
 Univ of Cologne,  Germany, preprint 1990 (submitted to
Phys. Rev. Lett.)

\newpage

\section*{Recent preprints}

\item
D. Armand-Ugon, R. Gambini, J. Griego, and L. Setaro.
Classical loop actions of gauge theories.
hep-th/9307179

\item A. Ashtekar.
Mathematical problems of non-perturbative quantum general relativity.
To appear in: Proceedings of the 1992 Les Houches summer school on gravitation
and quantization, Ed. B. Julia (North-Holland, Amsterdam, 1993)

\item
A. Ashtekar, R.S. Tate and C. Uggla.
Minisuperspaces: symmetries and quantization.
In {\em Misner Festschrift}, edited by B.L. Hu, M. Ryan and C.V.
Vishveshwara (Cambridge University Press, to appear 1993)

\item
J.C. Baez.
Link invariants, holonomy algebras and functional integration.
Riverside (Dec.\ 1992)

\item
J.C. Baez.
Diffeomorphism-invariant generalized measures on the space of
connections modulo gauge transformations.
hep-th/9305045

\item
F. Barbero Gonzalez and M. Varadarajan.
The phase space of 2+1 dimensional gravity in the Ashtekar
formulation.
gr-qc/9307006

\item
C. Di Bartolo, R. Gambini, J. Griego.
The extended loop group: an infinite dimensional manifold associated
with the loop space.
IFFI/93.01, gr-qc/9303010

\item
I. Bengtsson.
Neighbors of Einstein's equations --- some new results.
G\"oteborg preprint ITP92-35

\item
I. Bengtsson.
Strange reality: Ashtekar variables with variations
G\"oteborg preprint ITP92-36

\item
I. Bengtsson.
Form connections.
gr-qc/9305004

\item
Y. Bi and J. Gegenberg.
Loop variables in topological gravity.
gr-qc/9307031

\item
R. Brooks.
Diff($\Sigma$) and metrics from Hamiltonian-TQFT.
MIT preprint CTPH2175

\item
R. Capovilla.
Non-minimally coupled scalar field and Ashtekar variables.
To appear in Phys. Rev. D

\item
R. Capovilla, J. Dell and T. Jacobson.
The initial value problem in light of Ashtekar's variables.
UMDGR93-140, gr-qc/9302020

\item
S. Carlip.
Six ways to quantize (2+1)-dimensional gravity.
gr-qc/9305020

\item
Y.M. Cho, K.S. Soh, J.H. Yoon and Q.H. Park.
Gravitation as gauge theory of diffeomorphism group.
???

\item
G\'eza F\"ul\"op.
About a Super-Ashtekar-Renteln Ansatz.
gr-qc/9305001

\item
James Grant.
On self-dual gravity.
DAMTP-R92/47, gr-qc/9301014

\item
V. Husain.
Faraday lines and observables for the Einstein-Maxwell theory.
gr-qc/9306024

\item
H. Iwasaki and C. Rovelli.
Gravitons as embroidery on the weave.
Pittsburgh and Trento preprint (1992)

\item
H. Iwasaki and C. Rovelli.
{}From knot states to gravitons: the map $M$.
To appear in \ijmp{}

\item
Giorgio Immirzi.
The reality conditions for the new canonical variables of general
relativity.
To appear in \cqg{}

\item
T. Jacobson and J. Romano.
The spin-holonomy group in general relativity.
UMDGR-92-208

\item
H. Kodama.
Quantum gravity by the complex canonical formulation.
gr-qc/9211022, \ijmp{} to appear

\item
C. Kozameh, W. Lamberti, and E.T. Newman
Holonomy and the Einstein equations.
???

\item
K. Kuchar.
Canonical quantum gravity.
gr-qc/9304012

\item
H. Kunitomo and T. Saro.
The Ashtekar's formulation for canonical, $N=2$ supergravity.
OU-HET 167 (April 1992), to appear in \ijmp{D}

\item
Stephen Lau.
Canonical variables and quasilocal energy in general relativity.
gr-qc/9307026

\item
R. Loll, J. Mour\~ao, J. Tavares.
Complexification of gauge theories.
hep-th/9307142

\item
J. Louko.
Holomorphic quantum mechanics with a quadratic Hamiltonian constraint.
gr-qc/9305003, to appear in \pr{D}

\item
J. Louko and D. Marolf.
Solution space of 2+1 gravity on ${\bf R} \times T^2$ in Witten's
connection formulation.
gr-qc/9308018

\item
N. Manojlovi{\' c} and G. Mena Marugan.
Non-perturbative canonical quantization of minisuperspace models:
Bianchi types I and II.
gr-qc/9304041

\item
D. Marolf.
Loop representations for 2+1 gravity on a torus.
Syracuse SU-GP-93/3-1

\item
D. Marolf.
An illustration of 2+1 gravity loop transform troubles.
gr-qc/9305015

\item
H.-J.~Matschull.
Solutions to the Wheeler-DeWitt constraint of canonical gravity
coupled to scalar matter fields.
gr-qc/9305025

\item
H.-J.~Matschull and H. Nicolai.
Canonical quantum supergravity in three dimensions.
gr-qc/9306018

\item
M. Miller and L. Smolin.
A new discretization of classical and quantum general relativity.
gr-qc/9304005

\item
O. Obregon, J. Pullin, and M. Ryan.
Bianchi cosmologies: new variables and a hidden supersymmetry.
gr-qc/9308001

\item
Peter Peld\'an.
Ashtekar's variables for arbitrary gauge group.
G\"oteborg ITP 92-17, to appear in \pr{D}

\item
Peter Peld\'an.
Actions for gravity, with generalizations: a review.
gr-qc/9305011

\item
J. Pullin.
Knot theory and quantum gravity: a primer.
University of Utah preprint (Jan. 1993) UU-REL-93/1/9, hep-th/9301028

\item
A. Rendall.
Unique determination of an inner product by adjointness relations in
the algebra of quantum observables.
To appear in \cqg{}

\item
J. Romano.
On the constraint algebra of degenerate relativity.
gr-qc/9306034

\item
C. Rovelli.
A generally covariant quantum field theory.
Trento and Pittsburgh preprint (1992)

\item
C. Rovelli and L. Smolin.
The physical hamiltonian in nonperturbative quantum gravity.
gr-qc/9308002

\item
L. Smolin.
Diffeomorphism invariant observables in quantum gravity from a
dynamical theory of surfaces.
Syracuse preprint (1992), submitted to \np{B}

\item
L. Smolin.
What can we learn from the study of non-perturbative quantum general
relativity?
gr-qc/9211019

\item
L. Smolin.
Time, measurement and information loss in quantum cosmology.
gr-qc/9301016, to appear in the Brill festschrift

\item
L. Smolin
Finite, diffeomorphism invariant observables in quantum gravity.
SU-GP-93/1-1, gr-qc/9302011

\item
C. Soo and L. Chang.
Superspace dynamics and perturbations around ``emptiness''.
gr-qc/9307018

\item
J. Tavares.
Chen integrals, generalized loops and loop calculus.
Preprint, U. Porto (April 1993)

\item
T. Thiemann.
On the solutions to the initial value constraints for general
relativity for general matter coupling in terms of Ashtekar's variables.
Aachen PITHA 93-1


\end{enumerate}

\end{document}